\documentclass[twocolumn,showpacs,preprintnumbers,amsmath,amssymb]{revtex4}

\usepackage{graphicx}
\usepackage{dcolumn}
\usepackage{bm}

\begin{document}
\def\brho{{\hbox{\boldmath $\rho$}}}
\def\bsb{{\hbox{\boldmath $\beta$}}}
\def\bsk{{\hbox{\boldmath $k$}}}
\def\bsp{{\hbox{\boldmath $p$}}}

\title{Analysis of pion elliptic flows and HBT interferometry in a 
granular quark-gluon plasma droplet model} 

\author{Wei-Ning Zhang$^{1,2,3}$}
\author{Yan-Yu Ren$^2$}
\author{Cheuk-Yin Wong$^{3,4}$}

\affiliation{
$^1$Department of Physics, Dalian University of Technology, 
Dalian, Liaoning 116024, P. R. China\\
$^2$Department of Physics, Harbin Institute of Technology, 
Harbin, Heilongjiang 150006, P. R. China\\
$^3$Physics Division, Oak Ridge National Laboratory, Oak Ridge, TN
37831, U.S.A.\\
$^4$Department of Physics, University of Tennessee, Knoxville, TN
37996, U.S.A.
}

\date{\today}

\begin{abstract}
In many simulations of high-energy heavy-ion collisions on an
event-by-event analysis, it is known that the initial energy density
distribution in the transverse plane is highly fluctuating.
Subsequent longitudinal expansion will lead to many longitudinal tubes
of quark-gluon plasma which have tendencies to break up into many
spherical droplets because of sausage instabilities.  We are therefore
motivated to use a model of quark-gluon plasma granular droplets that
evolve hydrodynamically to investigate pion elliptic flows and
Hanbury-Brown-Twiss interferometry.  We find that the data of
pion transverse momentum spectra, elliptic flows, and HBT radii in
$\sqrt{s_{NN}}=200$ GeV Au + Au collisions at RHIC can be described
well by an expanding source of granular droplets with an anisotropic
velocity distribution.
\end{abstract}

\pacs{25.75.-q, 25.75.Nq, 25.75.Gz}

\maketitle

\section{Introduction}
                                                                                
Recently, there has been much progress in our understanding of the
process of nucleus-nucleus collisions at RHIC
\cite{Gyu04,Rik04,BRA05,PHO05,STA05,PHE05}.  Hydrodynamical
calculations agree well with the RHIC $v_2$ data of the elliptic flow
at low transverse momentum $p_{_T} < 2$ GeV
\cite{STA01,STA02,STA03,PHE03,STA04}.  However, they cannot predict
the saturation of $v_2$ at higher $p_{_T}$
\cite{STA02,STA03,PHE03,STA04} and the RHIC HBT puzzle of $R_{\rm
out}/R_{\rm side} \approx 1$ \cite{STA01a,PHE02a,PHE04a,STA05a}.  The
HBT puzzle is contrary to many earlier theoretical expectations
\cite{Ris96,Tea99,Wie99,Sof01,Wei02,Pra03}.  Various models have been
put forth to explain the HBT puzzle
\cite{Sof02,Hei02,Lin02,Tea03,Cso03,Mol04,Zha04,Soc04,Won03,Cra05,Lis05,Pra05,Fro06}.
There have also been many attempts to provide a consistent explanation
for both the elliptic flow and HBT measurements at RHIC
\cite{Hir02,Huo03,Hum03,Csa04,Lin05,Tom05,Ham05}.

In Ref.\ \cite{Zha04}, a granular particle-emitting source of
quark-gluon plasma (QGP) droplets evolving hydrodynamically was put
forth to explain the RHIC HBT puzzle.  The suggestion was based on the
observation that in the hydrodynamical model \cite{Ris96,Ris98}, the
particle emission time scales with the initial radius of the droplet.
Particles will be emitted earlier if the radius of the droplet is
smaller, as in a source of many droplets.  An earlier emission time
will lead to a smaller extracted HBT radius $R_{\rm out}$, while the
extracted HBT radius $R_{\rm side}$ is determined by the scale of the
distribution of the droplets.  As a result, the value of $R_{\rm
out}$ can be close to $R_{\rm side}$ for a granular quark-gluon plasma
source \cite{Zha04}.  Further suggestions of using the single-event
intensity interferometry to search for the signature for granular
structures have also been presented \cite{Won04,Zha05}.

Motivated by the successes of our previous analysis of the HBT puzzle
using a granular droplet model, we examine in this paper the
theoretical basis for the possible occurrence of granular structure in
the evolution of a quark-gluon plasma.  In addition, we wish to refine
the granular source model of Ref.\ \cite{Zha04} by considering more
reasonable anisotropic velocity distributions of the droplets instead
of a constant radial velocity assumed in \cite{Zha04}.  We would like
to investigate the pion elliptic flows as well as HBT radii as
functions of the pion transverse momentum.  The agreement both of the
$v_2$ and HBT radii with experimental data concurrently will gives
strong constrains for the granular model and its parameters and will
provide a useful insight into the initial state and the evolution of
the particle-emitting source produced in high-energy nucleus-nucleus
collisions.

\section{Granular instability in the Evolution of the 
quark-gluon plasma}

Based on the recent results of high-energy heavy-ion collisions at
RHIC, the matter (presumably QGP) produced in the collisions may be a
strongly-coupled medium with a very high energy density
\cite{Gyu04,Rik04,BRA05,PHO05,STA05,PHE05}.  They are thermalized
within about 1 fm/c \cite{Gyu04,Rik04,BRA05,PHO05,STA05,PHE05}.  It is
of interest to study its subsequent evolution and see whether granular
structures may play a role in the space-time development of the
presumed quark-gluon plasma matter.

Although a granular structure was suggested earlier as the signature
of a first-order phase transition \cite{Wit84}, the occurrence of
granular structure may not be limited to the occurrence of a
first-order phase transition. There are additional effects which may
lead to the dynamical formation of granular droplets.

Because the QGP is a strongly interacting dense medium, a surface
tension arises at a boundary due to the presence of a strong
interaction in the dense phase on one side of the boundary and the
absence (or weakening) of the strong interaction with no density (or
diminishing density) on the other side
\cite{Mye74,Mol93,Kaj86,Iwa92,Gro93,Bib97,Cse92,Kis05}.  This
imbalance of the forces acting on different parts at the boundary
leads to the surface tension, while the detail profile of the boundary
may depend on the nature and the order of phase transition.  Due to
the presence of this surface tension, there may be instability against
surface shape changes and bulk density oscillations.

\subsection{Transverse Density Fluctuations and Surface Granular Instability} 

The quark gluon plasma that may be produced in the laboratory is
spatially bounded by various boundaries.  Furthermore, in many
simulations of the heavy-ion collisions on an event-by-event basis,
the initial transverse energy density is far from being uniform.  It
exhibits large transverse density fluctuations with a large
peak-to-valley ratio \cite{Dre02,Ham05,Soc04}.  A large number of
transverse density domains are clearly visible in the initial
transverse density distribution of the produced matter in Fig. 21 of
\cite{Dre02} or Fig. 4 of \cite{Ham05}.  Because of these highly
fluctuating initial transverse density distribution forming ``lumps''
in different regions, the probability for the occurrence of granular
structure may be enhanced.

We can present an approximate scenario to discuss the evolution of
matter with a large transverse density fluctuations, in a central
collision of two heavy and equal nuclei of radius $R$.  Because of
Lorentz contraction, the longitudinal length $R/\gamma$ of the
produced matter is much smaller than the transverse length, which is
of the order $R$.  Initially as depicted in Fig. 1(a), density
fluctuations in the transverse plane manifest as a number of
transverse lumps characterized by a radius $r_{d0}$, which one can also
see from the results of \cite{Dre02} and \cite{Ham05}.  From such an
initial data, we can consider the evolution according to Landau
hydrodynamics \cite{Lan53}, which gives a good description of the
longitudinal distribution of the produced particles
\cite{Ste05,Mur04}.  In such a description, because the initial
longitudinal length is much smaller than the transverse length, the
hydrodynamical force per unit volume in the longitudinal direction,
$-\partial P /\partial z$, is much greater than the hydrodynamical
force per unit volume in the transverse direction, $-\partial P
/\partial \rho$.  The subsequent expansion in the longitudinal
direction proceeds much faster than in the transverse direction, as
pointed out by Landau and Belenkii \cite{Lan53}.  The expanding matter
develops into tubes of radius $r_{d0}$ [Fig.\ 1($b$)] and the density of
matter in the tube decreases as a function of the proper time.

\begin{figure}
\includegraphics[angle=0,scale=0.40]{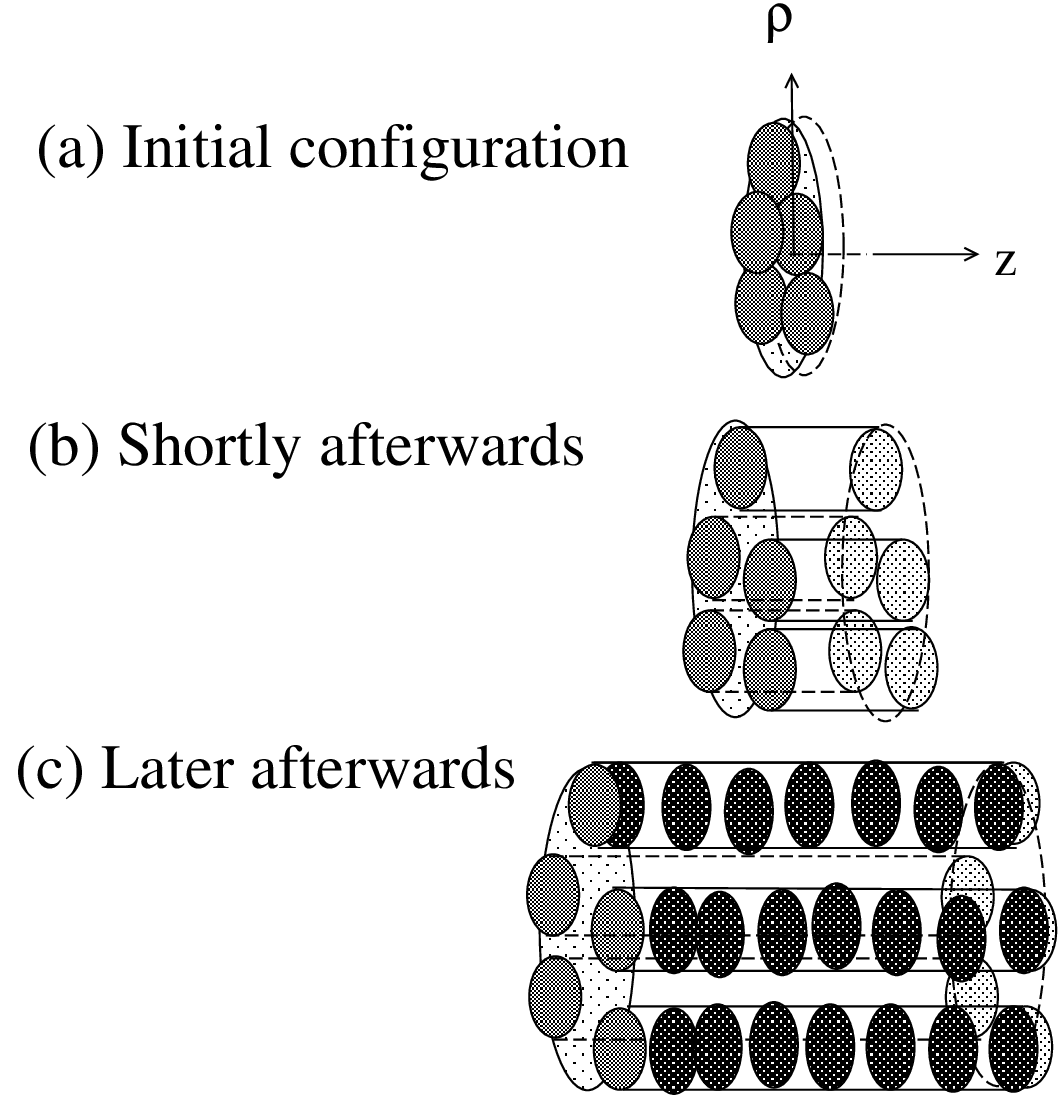}
\caption{\label{fig:zexpnlg} Schematic description of the matter
 distribution under a rapid longitudinal expansion in the
 $z$-direction, for a highly fluctuating initial transverse density
 distribution [Fig.\ 1$a$].  The system evolves into many longitudinal
 tubes at Fig.\ 1$b$ and later at Fig.\ 1$c$.  Sausage instabilities
 lead to the production of droplets along the longitudinal tubes as
 depicted in Fig. 1$c$.  }
\end{figure}

As is well known, a long tube of matter is unstable against
``sausage'' perturbations that tend to make the tube narrow in some
regions and thicker in other regions. The sausage instability leads to
the break-up of the long tube of matter into many approximately equal
spherical droplets, as in the break-up of toroidal liquid droplets
\cite{Won73}.

We can examine the sausage instability for a cylinder with a length
$L$ and a radius $r_{d0}$, with $L>> r_{d0}$.  
We consider ``sausage'' perturbations $a_n$ about the cylinder of matter,
\begin{eqnarray}
r_d(z)=\tilde r_{d0}
\left \{ 1 + \sum_{n={\rm odd}}a_n \sin(n\pi z/L)\right \}.
\end{eqnarray}
In order to have a positive slope $dr_d(z)/dz$ at $z=0$ and a negative
slope at $z=L$, we need to limit ourselves to perturbations with odd
values of $n$ and $a_n>0$.  By the condition of volume conservation,
the quantity $\tilde r_{d0}$  is 
\begin{eqnarray}
\tilde r_{d0}=r_{d0}\Biggl  \{ 1 - \!\! \sum_{n={\rm odd}} \! \frac {2a_n}{n\pi} 
- \frac {1}{4} \! \sum_{n={\rm odd}} \! {a_n^2}
+ \frac {3}{2} \bigg( \sum_{n={\rm odd}}\frac {2a_n}{n\pi} \bigg)^{\!\!2}
\Biggr \}.
\end{eqnarray}
The surface energy of the cylinder is then
\begin{eqnarray}
E_s=\sigma 2 \pi r_{d0} L 
\Biggl  \{ 1 - \frac {1}{4}  \sum_{n={\rm odd}}  {a_n}^2
+ \frac {1}{2} \bigg( \sum_{n={\rm odd}}\frac {2a_n}{n\pi} \bigg)^{\!\!2}
\Biggr \}.
\end{eqnarray}
The cylinder is unstable against perturbations of order $n$ if 
$d^2E_s/d a_n^2$ is negative.  We can evaluate $d^2E_s/d a_n^2$ and obtain
\begin{eqnarray}
d^2 E_s/d a_n^2 = \sigma 2 \pi  r_{d0}  L \{ -1/2 + (2/n \pi)^2 \}.
\end{eqnarray}
Thus, $ d^2 E_s/d a_n^2 $ is negative for all odd integer values of
$n$.  As an unrestrained growth of the perturbation of order $n$ leads
the formation of $(n+1)/2$ droplets, we find that the cylinder is
unstable against perturbations leading to the formation of $1,2,3,...$
approximately equal droplets.  Thus, when the length $L$ of the tubes
are stretched, sausage (granular) instability will develop and the
long tube will fragment into approximately equal-size droplets, as
depicted in Fig. 2$c$.  Large density fluctuations in the transverse
direction, together with the surface tension effects, favor the
formation of granular droplets.

In our simplified description, we have ignored the effects of the
perturbation on the variation of the transverse surfaces at the two
ends of the cylinder.  A refinement on the description of the end
surfaces will not modify significantly the question of the instability
of the cylindrical matter, for cylinders with $L$ substantially
greater than $r_{d0}$.

In conclusion, we find from the above analysis that the occurrence of
granular droplets may be more common than previously thought.  It is
of great interest to explore the subsequent dynamics in a granular
model for the description of various observables measured in
high-energy heavy-ion collisions.  This is particularly relevant
because a granular droplet model can explain the occurrence of early
emission of particles and a small value of $R_{\rm out}$ \cite{Zha04}.

\subsection{Bulk density oscillations due to surface tension interactions}

To carry out the analysis of the effects of surface tension on the
stability of density oscillations, we consider the following simple
model of the surface tension, as represented by a finite-range
interaction.  The equation of hydrodynamics without a finite-range
interaction is \cite{Lan59}
\begin{eqnarray}
\label{eq1}
[\epsilon + p] u^k \frac{\partial u_i}{\partial x^k}
= -   \frac{\partial p}{\partial x^i}
  -   u_i u^k  \frac{\partial p}{\partial x^k} \,,
\end{eqnarray}
where $\epsilon$ and $p$ are the energy density and pressure.  In this
hydrodynamical equation, interactions between constituents leading to
bulk properties of the plasma have already been included into the
characterization of the pressure $p$ and the energy density
$\epsilon$.  What remains is the effective residual finite-range
interaction $v({\bf r}-{\bf r}')$ that can lead to the surface
tension.  In the presence of this finite-range residual interaction,
the above can be generalized to be \cite{Won76}
\begin{eqnarray}
[\epsilon({\bf r}) &+& p({\bf r})]
u^k({\bf r})  \frac{\partial u_i({\bf r})}{\partial x^k}
= -   \frac{\partial p({\bf r})}{\partial x^i}
  -   u_i({\bf r}) u^k({\bf r})  \frac{\partial p({\bf r})}{\partial x^k} 
\nonumber \\
&-& \int d{\bf r'} \frac{\epsilon ({\bf r})}{W}
\frac{ \epsilon ({\bf r}')}{W} \nabla_i v({\bf r}-{\bf r}') \,,
\end{eqnarray}
where $W$ is the average energy per particle which may be
temperature-dependent, and $\epsilon({\bf r})/W$ gives the number
density $n({\bf r})$ of the constituents at ${\bf r}$.

We choose to represent the effect of the surface tension in terms of an
effective residual interaction of the following form as used in
\cite{Mol93},
\begin{eqnarray}
\label{surf}
v(r)=\alpha_r (1-\frac{\mu r}{2}) \frac {e^{-\mu r}}{r} \,,
\end{eqnarray}
where $\alpha_r>0$ and $1/\mu$ are the strength and the range of the
residual interaction that leads to the surface tension.  As a residual
interaction, $v(r)$ has been chosen such that the contribution of the
interaction to the total energy of a system with a uniform density is
zero and is positive and proportional to the surface area in the
surface region.  

To carry out the stability analysis, we can envisage that by solving
the hydrodynamical equation (\ref{eq1}) without the surface
interaction, one obtains a comparatively slow-varying density
distribution $\epsilon_0({\bf r},t)$ with a local average $\bar
\epsilon_0= \langle \epsilon_0({\bf r},t) \rangle$.  Then following
the derivations of the equation of sound waves \cite{Lan59}, the
perturbation $\delta \epsilon({\bf r},t)= \epsilon({\bf r},t) -
\epsilon_0({\bf r},t)$ obeys the wave equation
\begin{eqnarray}
\label{eq4}
\frac {\partial^2 \delta \epsilon ({\bf r,t})} {\partial t^2}
& \approx & c_s^2 \nabla^2 \delta \epsilon ({\bf r,t})
+ \nabla_{\bf r} \cdot \int d{\bf r'}
[ (\delta  \epsilon({\bf r},t) \epsilon({\bf r}',t) \nonumber \\
&& + \,\epsilon({\bf r},t) (\delta \epsilon({\bf r}',t))]
\nabla_{\bf r}   v({\bf r}-{\bf r}')/W^2 \,,
\end{eqnarray}
where $c_s$ is the speed of sound.  We consider density perturbations
of the type
\begin{eqnarray}
\delta \epsilon ({\bf r},t)
=e^{ i{\bf k} \cdot {\bf r}-i\omega t} \,.
\end{eqnarray}
Representing the spatial
variation of $\epsilon_0({\bf r},t)$ by its average
$\bar \epsilon_0$, we obtain from Eq.\ (\ref{eq4})
\begin{eqnarray}
\label{omega}
\omega^2=c_s^2 f({\bf k}) {\bf k}^2 \,,
\end{eqnarray}
for points in the interior of the medium. 
Here, $f({\bf k})$ is
\begin{eqnarray}
\label{fk}
f({\bf k})= 1-\frac {4 a \mu^2 {\bf k}^2 }
                  { ({\bf k}^2 + \mu^2)^2} \,,
\end{eqnarray}
where the second term of $f({\bf k})$ arises from the surface tension
interaction $v(r)$, and $a$ is the granular stability number defined as
\begin{eqnarray}
\label{eqa}
a= \frac { {\bar \epsilon_0}  \pi \alpha_r}{W^2 \mu^2 c_s^2} \,.
\end{eqnarray}

\begin{figure}
\includegraphics[angle=0,scale=0.40]{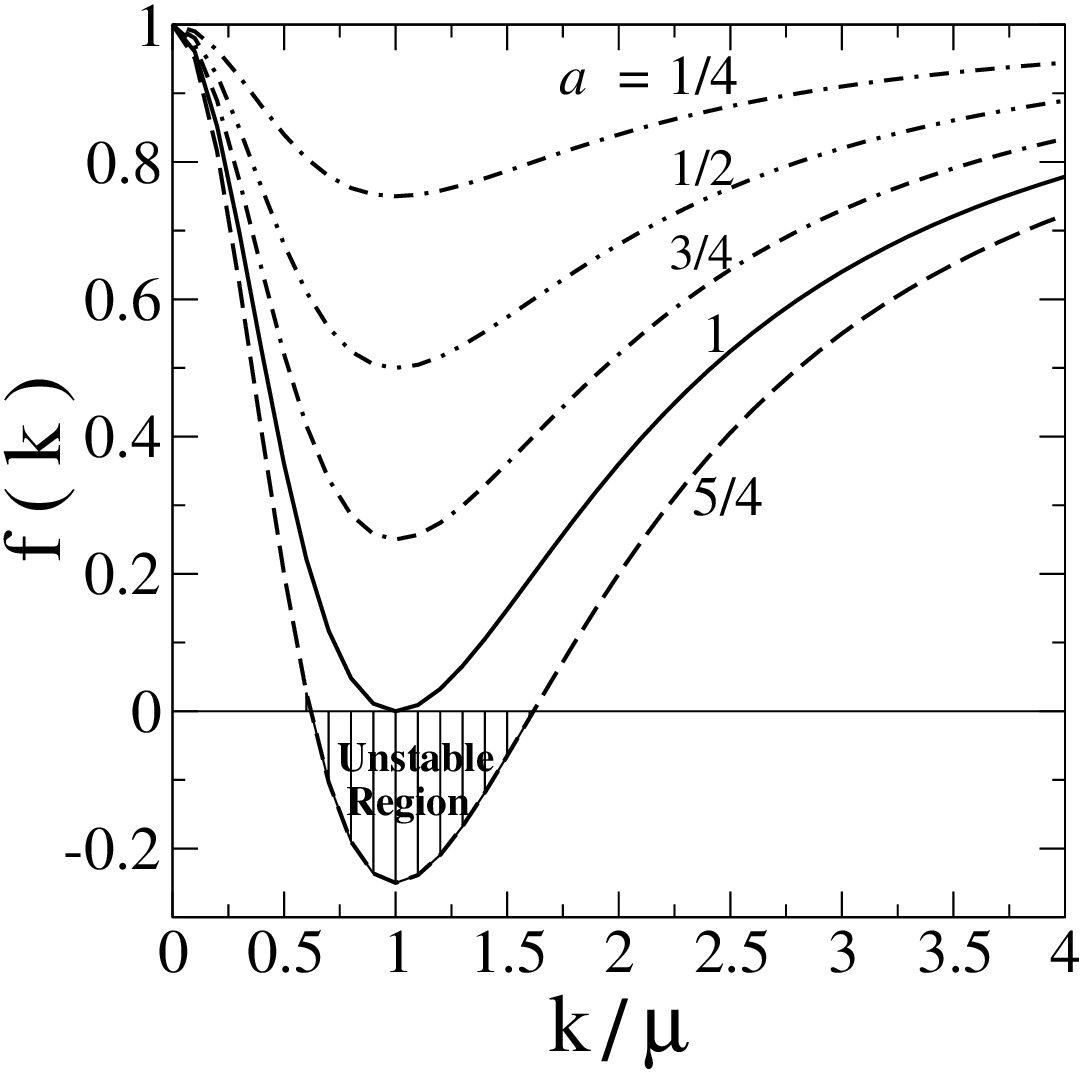}
\caption{\label{fig:fk} $f({\bf k})$ curves for different values of
$a$.  The expansion is unstable in the shadowy region.  }
\end{figure}

Equations (\ref{omega}) and (\ref{fk}) indicates that when $f(k)$
becomes negative, the system is unstable against a density
perturbation of wave number $k$. We plot in Fig.\ 2 the function
$f(k)$ as a function of $k/\mu$.  As one observes, the minimum of $f({
k})$ is located at $k/\mu=1$ for which the characteristic wave length
of the oscillation $1/k$ is equal to the range of the interaction
$1/\mu$.  The function $f(k)$ becomes negative when $a \ge 1 $ and it
is therefore appropriate to call $a$ the granular stability number.
When the granular stability number $a$ is greater than unity, the
system is unstable against perturbations with wave numbers $k$ in the
interval $k_-\le k \le k_+$, with $k_\pm$ given by
\begin{eqnarray}
{\bf k}_{\pm}^2 = \left \{ 2a - 1 \pm \sqrt{(2a-1)^2 - 1} \right \} \mu^2 \,.
\end{eqnarray}

We note from Eq.\ (\ref{eqa}) that the condition $a \ge 1$ for the
occurrence of granular instability happens when the surface tension
interaction strength $\alpha_r$ is large.  It will be useful to relate
$\alpha_r$ to the surface tension coefficient $\sigma$.  For such a
purpose, we study the energy of a system with a numbers density
profile given by
\begin{eqnarray}
n({\bf r})=n_{\rm out} + \frac{n_0-n_{\rm out}}{1+e^{z/a}},
\end{eqnarray}
where $n_0 (> n_{\rm out})$ is the number density in the interior
dense region of a surface and $n_{\rm out}$ is the number density in
the outer dilute region.  The energy of the system due to the
residual interaction of Eq.\ (\ref{surf}) is
\begin{eqnarray}
\label{eq5}
E=\frac{1}{2} \int d{\brho}\, dz \,  n({\bf r}) 
\int d{\brho}'\, dz'\,  n({\bf r}')  v({\bf r}-{\bf r}').
\end{eqnarray}
We can introduce $V(z)$ as
\begin{eqnarray}
V(z) = \int d{\brho}'dz'  n({\bf r}')  v({\bf r}-{\bf r}').
\end{eqnarray}
After integration the ${\brho}'$ coordinates, we find
\begin{eqnarray}
\label{eq7}
V(z)&=& \frac {\pi \alpha_r}{\mu}   \int_{-\infty}^{\infty} 
dz'\, n(z') \nonumber\\
&\times&\left \{ e^{-\mu|z-z'|}
-\mu|z-z'| e^{-\mu|z-z'|} \right \}.
\end{eqnarray}
From Eqs.\ (\ref{eq5})-(\ref{eq7}), the surface tension coefficient is
therefore given by
\begin{eqnarray}
\sigma &\equiv & dE/d{\brho}\nonumber  \\
&=&
\frac { \alpha_r \pi}{2\mu}
\int_{-\infty}^{\infty} dz\,  n(z) \int_{-\infty}^{\infty} dz' n(z')
\nonumber \\
&\times&
\left \{ e^{-\mu|z-z'|}
-\mu|z-z'| e^{-\mu|z-z'|} \right \},
\end{eqnarray}
which gives a relation between $\sigma$ and $\alpha_r$ for a general
density profile.  For the simple case of a sharp boundary with
$n(z)=n_0\theta(-z)$, the above equation gives the relationship
\begin{eqnarray}
\label{sig}
\sigma= \frac {n_0^2 \pi \alpha_r}{2 \mu^3}.
\end{eqnarray}
The surface tension coefficient increases with $\alpha_r/\mu^3$ and
the square of the number density.  As $n_0$ varies with
temperature as $T^3$ while $\alpha_r$ and $\mu$ are relatively slowly
varying functions of $T$, the surface tension varies with temperature
approximately as $T^6$.  When the surface is described by a sharp
surface profile, the granular stability number $a$ is related to the
surface tension coefficient $\sigma$ by
\begin{eqnarray}
a=\frac {2\mu \sigma}{\epsilon_0 c_s^2}.
\end{eqnarray}

In quenched QCD lattice gauge calculations, the surface tension
coefficient $\sigma(T_c)$ has been calculated for $T=T_c$, and
$\sigma(T_c)/T_c^3$ was found to range from 0.016 to 0.092
\cite{Iwa92,Gro93,Bib97,Kis05}.  If we use such a value of the surface
tension coefficient to evaluate the granular stability number $a$
using Eq.\ (\ref{eqa}), the quantity $a$ will have the value from
0.024 to 0.140 and the system is stable against density oscillations
of the type discussed here.  However, the medium considered by the
quenched QCD approximation consists of gluons without dynamical
quarks.  As is well known, the presence of dynamical quarks has
important influences on many thermodynamical properties of the plasma.
The phase transition temperature is altered from 269 MeV for quenched
QCD without dynamical quarks to 154 MeV for full QCD with 3 flavors of
quarks \cite{Kar01}, whereas the surface tension coefficient $\sigma$
depends on the temperature approximately as $T^6$.  The number density
of a quark-gluon plasma with three flavors of quarks is about two
times the number density of the plasma without dynamical quarks, while
the surface tension coefficient varies as the square of the number
density.  The surface tension coefficient $\sigma$ calculated in
quenched QCD at $T=T_c$ may be quite different from that calculated in
full QCD at higher temperatures.  It will be of interest to evaluate
the surface tension coefficient in full QCD and study its temperature
dependence in order to find out whether the quark-gluon plasma is
stable against bulk density oscillations with wave lengths close the
the range of the interaction discussed here.

The above discussions was carried out in the linearized perturbative
theory of slow hydrodynamical motion.  We envisage that the initial
QGP matter is highly compressed in the longitudinal direction and the
subsequent longitudinal expansion will be rather rapid and non-linear
in the density changes.  In this respect, it is of interest to note
that in previous calculations in full hydrodynamics for the rapid
expansion of a nuclear system with a finite range interaction, the
one-dimensional fragmentation of the density into one-dimensional
``lumps' occur, as shown in the left panel of Fig. 2 of
Ref. \cite{Won77}.  If these hydrodynamical calculations can be a
useful analogue, the analogous occurrence of granular droplets in the
bulk expansion of the quark-gluon plasma may be possible.

\section{Model of Granular QGP droplets}

Based on the above discussions, we would like to use a model of
granular quark-gluon plasma droplets to examine the dynamical
evolution of the system.  We assume that the system consists of $N_d$
droplets of radius $r_d$ initially distributed in a short cylinder of
length $2{\cal R}_z$ along the beam direction ($z$ direction) as in
Fig.\ 2$c$, with an initial transverse spatial distribution up to a
radius ${\cal R}_t$,
\begin{eqnarray}
\label{distri}
\frac{dP_d}{2\pi \rho \,d\rho \,dz} &\propto& [1-\exp(-\rho^2 /
\Delta{\cal R}_t^2)] \nonumber \\
& & \times \,\theta({\cal R}_t -\rho) \,\theta({\cal R}_z -|z|) \,,
\end{eqnarray}
where ${\rho}=\sqrt{x^2+y^2}$ and $z$ are the coordinates of the
center of a droplet.  The parameter $\Delta {\cal R}_t$ describes a
shell-type radial distribution which may arise from the dynamics of
the transverse expansion.  A naive blast-wave-type expansion with a
constant radial velocity would leave a void along the central
transverse axis at $\rho=0$.  
Instead of a constant transverse velocity,
we shall find later that the best parameters that describe the
experimental data suggests a transverse velocity profile
$({\beta_d})_T \sim \rho^{0.42}$.  From the equation of continuity,
the density would change approximately by
\begin{eqnarray}
\frac{dn}{dt} \sim - n \frac {1}{\rho} \frac {d} {d\rho} 
\rho (\beta_d)_{_T} \propto - 1.42\frac {n}{\rho^{0.58}}.
\end{eqnarray}
Thus the density decreases more rapidly near $\rho\sim 0$ and there is
a tendency to form a shell-type radial distribution.  The shell-type
distribution of the droplets may also arise from the shadowing of
detected particles originating from the central region of $\rho\sim0$.
The HBT radii as a function of the average pion transverse momentum of
a pion pair turns out to be slightly sensitive to this parameter of
$\Delta {\cal R}_t$.

Because of the early thermalization and the anisotropic pressure
gradient, the droplets will acquire anisotropic initial velocities.
We assume that the velocity of a droplet depends on the initial
coordinates of the droplet center, $(r_1, r_2, r_3)=(x,y,z)$, in the
form
\begin{eqnarray}
\label{bet}
(\bsb_d)_i = a_i \,{\rm sign}(r_i) \bigg(\frac{|r_i|}{{\cal R}_i} 
\bigg )^{\!\!b_i} \,, ~~~~~~~~~i=1,2,{\rm ~and~}3,
\end{eqnarray}
where $a_i$ is describes the magnitude of the anisotropic expansion,
${\rm sign}(r_i)$ denotes the sign of $r_i$, and $b_i\,(b_{_T},\,b_z)$
are the exponential power parameter that describes the variation of the
velocity with $r_i$.

For each single droplet, we use relativistic hydrodynamics and the
equation of state of the entropy density
\cite{Bla87,Lae96,Ris96,Ris98} to describe its subsequently evolution
\cite{Zha04,Zha05}.  The evolution of the granular source of many
droplets can then be obtained by superposing all of the evolutions of
individual droplets \cite{Zha04,Zha05}.  In our calculations, the
transition temperature and the transition temperature width are taken
to be $T_c=165$ MeV and $\Delta T=0.05T_c$
\cite{Ris96,Ris98,Zha04,Zha05}.  The initial energy density of the
droplets is taken to be $\epsilon_0 =3.75 T_c s_c$, which is about two
times of the density of quark matter at $T_c$ \cite{Ris96,Ris98}.  The
pions are emitted out of the surfaces of droplets at the freeze-out
temperature $T_f =0.95 T_c$, with momenta obeying the Bose-Einstein
distribution in the local frame at the temperature $T_f$.

In our model calculations there are five velocity parameters, $\{a_x$,
$a_y$, $a_z$, $b_{_T}=b_x=b_y$, and $b_z \}$, and four size parameters
of the granular source, $\{ {\cal R}_t$, ${\cal R}_z$, $\Delta{\cal
R}_t$, and $r_d\}$.  We find that the pion transverse momentum spectra
and the elliptic flow $v_2$ are sensitive to the velocity parameters and
insensitive to the size parameters.  While our HBT results are
sensitive both to the velocity parameters and the size parameters.  We
first determine the velocity parameters by comparing the pion
transverse momentum spectra and the elliptic flow $v_2$ of our model
calculations with experimental data.  Then, we determine the source
size parameters by comparing our results of HBT radii with
experimental data.

\begin{figure}
\includegraphics[angle=0,scale=0.65]{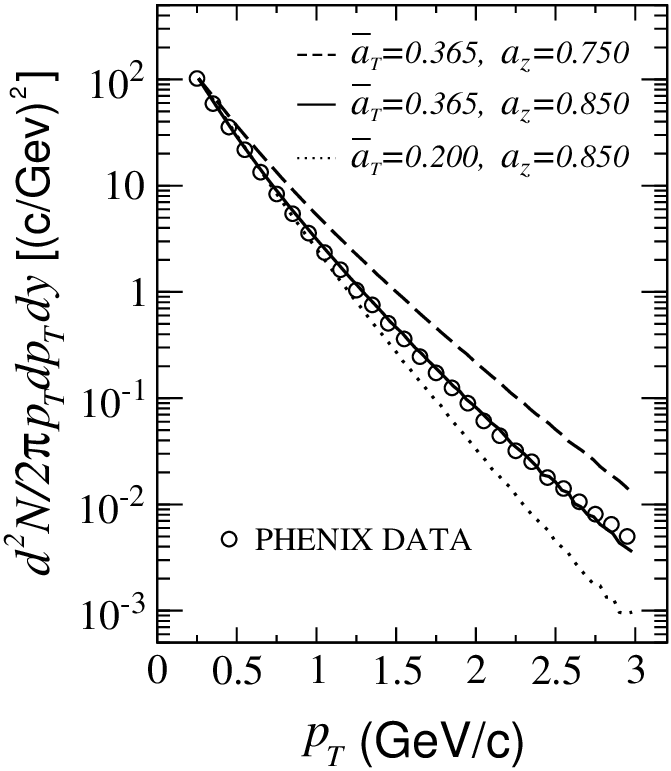}
\caption{\label{fig:f1} Negative pion transverse momentum spectrum for
$|y| < 0.5$.  The open circles give the PHENIX data for
$\sqrt{s_{NN}}=200$ GeV Au + Au collisions with minimum bias
\cite{PHE04b}.  Various curves give theoretical predictions with
different parameters.}
\end{figure}

Figure 3 shows the PHENIX $\pi^-$ transverse momentum spectrum in
$\sqrt{s_{NN}}=200$ GeV Au + Au collisions with minimum bias
\cite{PHE04b} and the transverse momentum spectrum obtained in our
model calculations.  The theoretical results have been normalized by
matching the distribution to the first $p_{_T}$ bin.  In our
calculations, we use a cut of particle rapidity $|y|<0.5$ in order to
compare with experimental data, which have been normalized to unit
rapidity \cite{PHE04b}.  Instead of the velocity magnitude parameters
of $a_x$ and $a_y$, we can equivalently use their average ${\bar
a}_{_T}=(a_x+a_y)/2$ and their difference $\Delta a_{_T}=a_x-a_y$.
The transverse momentum spectra depend on ${\bar a}_{_T}$ and is
independent of $\Delta a_{_T}$.  They also depend on $a_z$, $b_{_T}$,
and $b_z$.  We shall show that the elliptic flow $v_2$ is very
sensitive to $\Delta a_T$, $b_{_T}$ and $b_z$.  Good agreement with
experimental data of transverse momentum spectra and the elliptic flow can
be obtained concurrently by taking $b_{_T}=0.42$, $b_z=0.03$ and
$\Delta a_T=0.10$.  After $b_{_T}$ and $b_z$ are fixed, the parameters
${\bar a}_{_T}$ and $a_z$ can be obtained by comparing the transverse
momentum spectrum of the granular sources with experimental data
\cite{PHE04b}.  We finally determine ${\bar a}_{_T}=0.365$ and
$a_z=0.850$ for our granular source.

\section{Results of the elliptic flow}

The transverse momentum distribution of particles can be represented
in the form \cite{Won82,Oll92,Vol96}
\begin{eqnarray}
\label{ptd}
\frac{d^2N}{dp_{\!_T}^2d\phi}=\frac{dN}{2\pi dp_{\!_T}^2}\bigg[1
+2\sum_n v_n \cos(n\phi)\bigg]\,,
\end{eqnarray}
where $p_{_T}=\sqrt{p_x^2 +p_y^2}$ is the transverse momentum of the
particle, $\phi$ is its azimuthal angle with respect to the reaction
plane, and the harmonic coefficients, $v_n$ are anisotropy parameters.
The elliptic flow is defined as the second harmonic coefficient $v_2$,
which describes the eccentricity of the particle distribution in the
momentum space.  We choose the direction of $x$-axis in the reaction
plane and the direction of $y$-axis out of the reaction plane. We can
express $v_2$ as
\begin{eqnarray}
\label{v2d}
v_2 \equiv \langle \cos 2\phi \rangle =\bigg \langle \frac{p_x^2 -p_y^2}
{p_{\!_T}^2}\bigg \rangle\,.
\end{eqnarray}

\begin{figure}
\includegraphics[angle=0,scale=0.58]{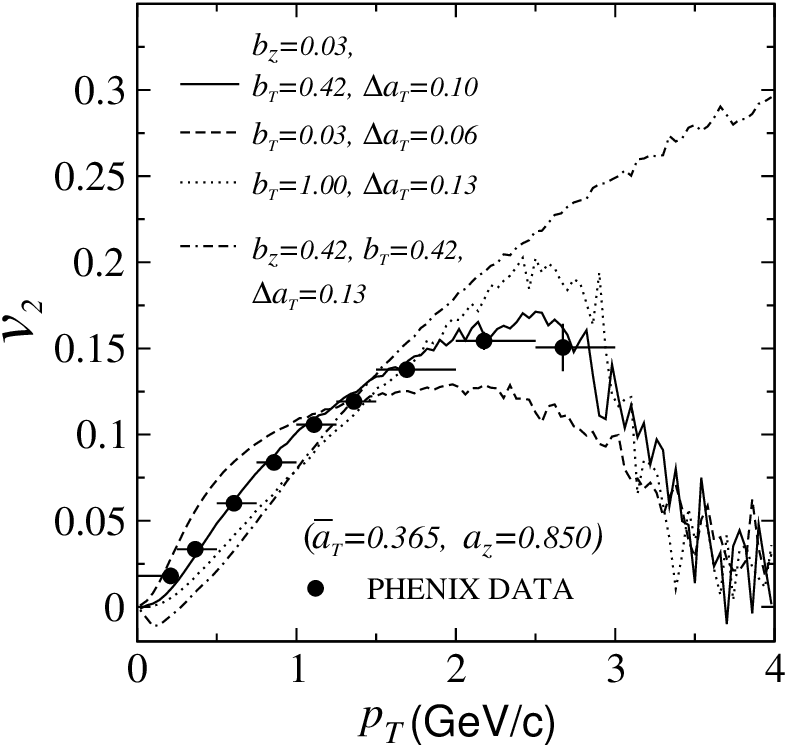}
\caption{\label{fig:f2} Pion elliptic flow $v_2$ as a function of particle 
transverse momentum for the granular sources and the experimental data of 
charged pions of PHENIX \cite{PHE03}.}
\end{figure}

Figure 4 shows the PHENIX $v_2$ data as a function of $p_T$ for
charged pions in Au + Au collisions at $\sqrt{s_{NN}}=200$ GeV
\cite{PHE03}.  Also shown are the theoretical $v_2$ results in our
granular droplet model. We use the same cut for particle
pseudo-rapidity, $|\eta|<0.35$, in theoretical calculations, as in the
experiment set-up \cite{PHE03}.  In the region $p_{_T}<2$ GeV, the
$v_2$ coefficient for a granular source increases with $p_{_T}$ as in
previous hydrodynamical calculations
\cite{Hir02,Huo03,Kol00,Kol01,Huo01,Kol03,Huo03a}.  However, in the high
$p_{_T}$ region with $p_{_T}>2.5$ GeV, we find that the $v_2$
coefficient for a granular source with small $b_z$ decreases with
$p_{_T}$, which is different from previous hydrodynamical calculations
\cite{STA02,STA03,PHE03,STA04,Huo03,Kol00,Kol01,Huo03a}.  In the
droplet-rest frame, the average transverse momentum of particles in
the $x$ and $y$ directions, $\langle p'_x \rangle$ and $\langle p'_y
\rangle$, are the same.  In the laboratory frame, a elliptic flow
arises from the anisotropic collective transverse boost of the
droplet.  For a very small $b_z$, the longitudinal velocity of
droplets is almost constant in the source while the magnitude of the
longitudinal boost velocity $a_z$ is very large (0.85) in our model.
In this case, the longitudinal component of droplet velocity is
important and it does not boost the transverse momenta of the
particles.  The transverse velocity of the droplet is much smaller
than the longitudinal velocity of the droplet.  Its boost effect
becomes weaker for the particles with higher transverse momentum
$p'_{_T}$ (higher $p_{_T}$).  On the other extreme for larger $b_z$,
the longitudinal velocity of droplet is small for the droplets with
small longitudinal coordinates.  The transverse components of droplet
velocity are dominant in the small $z$ region, which lead to a
production of the elliptic flow and the $v_2$ coefficient increasing
with transverse momentum \cite{Huo01}, even at higher $p_{_T}$ as
shown in the dashed-dot curve in Fig. 4.  

After ${\bar a}_{_T}$ and $a_z$ have been fixed at the values 0.365
and 0.850 as determined by the pion transverse momentum spectra, we
examine the pattern of $v_2$ as a function of transverse momentum
$p_T$ as we vary $b_{_T}$, $b_z$, and $\Delta a_{_T}$.  In the region
of $p_{_T}<2$ GeV, the curvature of $v_2$ is sensitive to $b_{_T}$.
For $b_z=0.03$, the curvature for $b_{_T} =0.03$ is too large (see the
dashed line in Fig. 4) and the curvature for $b_{_T}=1.00$ is too
small (see the dot line in Fig. 4).  For $b_z=b_{_T}=0.42$, the curve
of $v_2$ rises continuously and deviates from experimental data for
$p_{_{T}} > 1.5$ GeV.  We find that the set of parameters $\{b_z,
b_{_T}, a_{_T}\}= \{0.03, 0.42, 0.10\}$ give results consistent with
experimental $v_2$ data, as indicated by the solid theoretical curves
in Fig. 4.  The fact that $b_{_T}$ is much greater than $b_z$ while
$a_z$ is substantially greater than $a_T$ indicates that the dynamical
behavior in the transverse expansion and longitudinal expansion are
very different.

Recent experimental data of the elliptic flow of pion show that $v_2$
remains to have the value of 0.1 at very high $p_{_T}$ \cite{Sor05}.
The processes of particle production in the high $p_{_T}$ region are
dominated by parton and hard-probe processes and is beyond the thermal
emission model we consider here.

\section{Results of HBT interferometry}

The two-particle Bose-Einstein correlation function is defined as the
ratio of the two-particle momentum distribution $P(p_1,p_2)$ relative
to the the product of the single-particle momentum distribution
$P(p_1) P(p_2)$.  For a chaotic pion-emitting source, $P(p_i)~(i=1,2)$,
and $P(p_1,p_2)$ can be expressed as \cite{Won94}
\begin{eqnarray}
\label{Pp1}
P(p_i) = \sum_{X_i} A^2(p_i,X_i) \,,
\end{eqnarray}
\begin{eqnarray}
\label{Pp12}
P(p_1,p_2) = \sum_{X_1, X_2} \Big|\Phi(p_1, p_2; X_1,
X_2 )\Big|^2 ,
\end{eqnarray}
where $A(p_i,X_i)$ is the magnitude of the amplitude for emitting a
pion with 4-momentum $p_i=(\bsp_i,E_i)$ in the laboratory frame at
$X_i$ and is given by the Bose-Einstein distribution with freeze-out 
temperature $T_f$ in the local rest frame of the source point.  
$\Phi(p_1, p_2; X_1, X_2 )$ is the two-pion wave function.  Neglecting 
the absorption of the emitted pions by other droplets, $\Phi(p_1, p_2; 
X_1, X_2 )$ is simply
\begin{eqnarray}
\label{PHI}
& &\!\!\!\!\!\!\!\Phi(p_1, p_2; X_1, X_2 )
\nonumber \\
& & =\frac{1}{\sqrt{2}} \Big[ A(p_1, X_1)
A(p_2, X_2) e^{i p_1 \cdot X_1 + i p_2 \cdot X_2}
\nonumber \\
& & ~~~+ A(p_1, X_2) A(p_2, X_1)
e^{i p_1 \cdot X_2 + i p_2 \cdot X_1 } \Big] .~~~~~
\end{eqnarray}
Using the components of ``out", ``side", and ``long"
\cite{Pra90,Ber88} of the relative momentum of the two
pions, $q=|{\bf p_1}-{\bf p_2}|$, as variables, we can construct the
correlation function $C(q_{\rm out},q_{\rm side},q_{\rm long})$ from 
$P(p_1,p_2)$ and $P(p_1)P(p_2)$ by summing over ${\bf p_1}$ and ${\bf p}_2$ 
for each $(q_{\rm out},q_{\rm side},q_{\rm long})$ bin \cite{Zha04}.  The 
HBT radii $R_{\rm out}$, $R_{\rm side}$, and $R_{\rm long}$ can then be 
extracted by fitting the calculated correlation function $C(q_{\rm out},
q_{\rm side},q_{\rm long})$ with the following parametrized correlation 
function
\begin{equation}
\label{cq}
C(q_{\rm out},q_{\rm side},q_{\rm long})=1+\lambda \, e^{-q_{\rm out}^2 
R_{\rm out}^2 -q_{\rm side}^2 R_{\rm side}^2 -q_{\rm long}^2 R_{\rm long}^2} 
\,.
\end{equation}

\begin{figure}
\includegraphics[angle=0,scale=0.52]{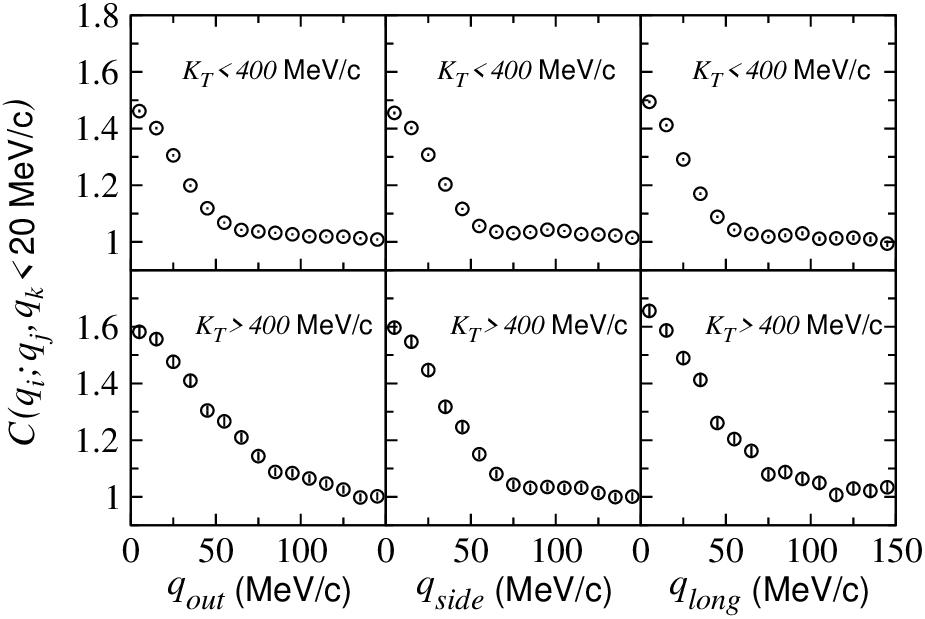}
\caption{\label{fig:f3} Two-pion correlation functions for granular source. 
}
\end{figure}

\begin{figure}
\includegraphics[angle=0,scale=0.6]{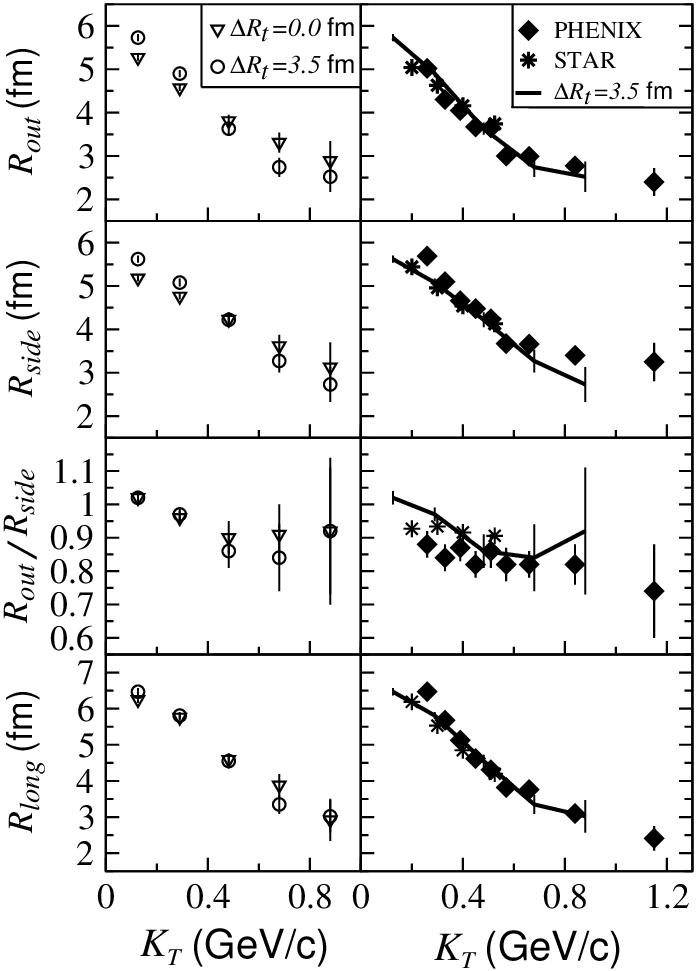}
\caption{\label{fig:f4} Two-pion HBT radii obtained by the PHENIX
Collaboration \cite{PHE04a} and the STAR Collaboration \cite{STA05}
compared with the theoretical results calculated in the granular
droplet model with $\Delta{\cal R}_t=0$ and 3.5 fm.  }
\end{figure}

Figure 5 shows the theoretical two-pion correlation functions for the
granular source.  The top figures give the results for a average pion
transverse momentum of a pion pair, $K_T$, less than 400 MeV/c, and
the bottom figures for $K_T > 400$ MeV/c.  The velocity parameters of
the granular source are ${\bar a}_{_T}=0.365$, $\Delta a_{_T}=0.10$,
$a_z=0.850$, $b_{_T}=0.42$, and $b_z=0.03$, which are determined by
the transverse momentum spectra and the elliptic flow $v_2$ as
discussed above.  The size parameters of the granular sources are
taken to be ${\cal R}_t=8.8$ fm, $\Delta{\cal R}_t=3.5$ fm, ${\cal
R}_z=7.0$ fm, and $r_d=1.3$ fm.  The number of droplet $N_d$ is taken
to be 40 in our calculations. One observes noticeable differences in
the lower $K_T$ region as compared to the higher $K_T$ region.

The left panels of figure 6 give the fitted two-pion HBT radii for the
granular source as a function of $K_T$.  The symbols of circle and
down-triangle are for $\Delta{\cal R}_t =0.35$ fm and $\Delta{\cal
R}_t =0$, respectively.  The parameter $\Delta{\cal R}_t$ of Eq.\
(\ref{distri}) describes the radial distribution of the centers of the
droplets, with $\Delta{\cal R}_t =0$ for a uniform distribution and
$\Delta{\cal R}_t =0.35$ fm for a shell-type radial distribution.  The
experimental PHENIX results \cite{PHE04a}, and STAR results
\cite{STA05} are shown on the right panels.  The curve gives the
theoretical results for $\Delta{\cal R}_t =0.35$ fm.  For our
theoretical HBT calculations, we use a cut for particle
pseudo-rapidity region $|\eta|< 0.35$, the same as in the PHENIX
experiments \cite{PHE04a}.  We find that if we increase the parameter
$r_d$, the HBT radii $R_{\rm out}$ and $R_{\rm long}$ will increase.
And if we increase the parameter $\Delta{\cal R}_t$, the variation of
HBT radii $R_{\rm out}$ and $R_{\rm side}$ with $K_T$ will become
steep.  The HBT results of granular source for $\Delta{\cal R}_t=3.5$
fm agree quite well with experimental data, although the case with
$\Delta {\cal R}_{_T}=0$ also give almost as good an agreement.

\section{Summary and Conclusion}

The expansion of the dense matter produced in high-energy heavy-ion
collisions may favor the production of granular droplets.  The strong
interaction associated with the dense matter will lead naturally to
the presence of a surface tension on a boundary.  

In an event-by-event basis, the initial transverse density
distribution of matter in central nucleus-nucleus collisions have been
known to be highly fluctuating \cite{Dre02,Ham05,Soc04}.  This large
spatial fluctuations may facilitate the formation of granular
droplets.  We envisage that as the expansion in the longitudinal
direction proceeds much more rapidly than in the transverse direction,
these initial transverse density fluctuations will lead to the
formation of many longitudinal tubes of matter with a length much
greater than its radius.  Due to the surface tension effects, these
tubes are unstable against granular sausage instability and will tend
to break up into approximately equal-size spherical droplets.

Motivated by the possibility for the occurrence of granular droplets
and our previous successful HBT analysis in terms of granular QGP
droplets, we refine our previous granular QGP droplet model to include
initial anisotropic velocities due to the anisotropic pressure
gradients at an earlier stage of the collisions.  We investigate
concurrently the data of (1) the pion transverse momentum spectra, (2)
the elliptic flow $v_2$ as a function of $p_{_T}$, and (3) various HBT
radii as a function of $K_{_T}$, for Au + Au collisions at
$\sqrt{s_{NN}}=200$ GeV at RHIC.  We find that the collection of
different pieces of experimental data can be described well by a
granular source model.  Direct confirmation of the granular droplet
configuration may need to await more experimental work on  single-event
HBT correlations and the fluctuation between correlation functions of
the single- and mixed-events, as suggested previously
\cite{Won04,Zha05}.

The extracted parameters from our analysis provide useful information
on the nature of the transverse and longitudinal expansion.  The
elliptic flow $v_2$ is due to the anisotropic initial dynamical
conditions in different transverse directions. Its decrease at high
$p_{_T}$ is sensitive to the exponential power parameter of the
droplet velocity $b_z$ in the longitudinal direction.  The value of
$b_z$ is much smaller than that of $b_{_T}$, indicating different
dynamical behaviors in longitudinal and transverse directions.  Our
HBT results further indicate that the granular source has a small
lifetime and a possible shell-type spatial distribution. Further
studies on the origin of the difference in the transverse and
longitudinal dynamics as well as the spatial distribution in the
transverse direction will be of great interest.

\begin{acknowledgments}
W.N.Z. would like to thank Dr. V. Cianciolo and Dr. G. Young for their kind 
hospitality at Oak Ridge National Laboratory.  We acknowledge discussions 
with Dr. T. Chujo, Dr. V. Cianciolo, Dr. B.-A. Li, and Dr. B. Zhang.  This 
research was supported by the National Natural Science Foundation of China 
under Contracts No. 10275015 and No. 10575024, and by the Division of Nuclear 
Physics, US DOE, under Contract No. DE-AC05-00OR22725 managed by UT-Battle, 
LC.
\end{acknowledgments}


\begin{thebibliography}{99}

\bibitem{Gyu04}
M. Gyulassy and L. McLerran, nucl-th/0405013.
                                                                                
\bibitem{Rik04} See also {\it New Discoveries at RHIC: the current
case for the strongly interactive QGP}, RIKEN Scientific Articles,
Volume 9, BNL, May 14-15, 2004.

\bibitem{BRA05}
BRAHMS Collaboration, I. Arsene {\it et al.}, Nucl. Phys. A {\bf 757}, 
1, 2005. 

\bibitem{PHO05}
PHOBOS Collaboration, B. B. Back {\it et al.}, Nucl. Phys. A {\bf 757}, 
28, 2005. 

\bibitem{STA05}
STAR Collaboration, J. Adams {\it et al.}, Nucl. Phys. A {\bf 757}, 
102, 2005. 

\bibitem{PHE05}
PHENIX Collaboration, K. Adcox {\it et al.}, Nucl. Phys. A {\bf 757}, 
184, 2005. 

\bibitem{STA01}
STAR Collaboration, C. Adler {\it et al.}, Phys. Rev. Lett. {\bf 87}, 
182301, 2001.

\bibitem{STA02}
STAR Collaboration, C. Adler {\it et al.}, Phys. Rev. Lett. {\bf 89}, 
132301, 2002.

\bibitem{STA03}
STAR Collaboration, C. Adler {\it et al.}, Phys. Rev. Lett. {\bf 90}, 
032301, 2003.

\bibitem{PHE03}
PHENIX Collaboration, S. S. Adler {\it et al.}, Phys. Rev. Lett. {\bf 91}, 
182301, 2003.

\bibitem{STA04}
STAR Collaboration, J. Adams {\it et al.}, Phys. Rev. Lett. {\bf 92}, 
052302, 2004.

\bibitem{STA01a}
STAR Collaboration, C. Adler {\it et al.}, Phys. Rev. Lett. {\bf 87}, 
082301, 2001.

\bibitem{PHE02a}
PHENIX Collaboration, K. Adcox {\it et al.}, Phys. Rev. Lett. {\bf 88}, 
192302, 2002.

\bibitem{PHE04a}
PHENIX Collaboration, S. S. Adler {\it et al.}, Phys. Rev. Lett. {\bf 93}, 
152302, 2004.

\bibitem{STA05a}
STAR Collaboration, J. Adams {\it et al.}, Phys. Rev. C {\bf 71}, 
044906, 2005.

\bibitem{Ris96}
D. H. Rischke and M. Gyulassy, Nucl. Phys A {\bf 608}, 479 (1996). 

\bibitem{Tea99}
D. Teaney and E. Shuryak, Phys. Rev. Lett. {\bf 83}, 4951 (1999). 

\bibitem{Wie99}
U. A. Wiedemann and U. Heinz, Phys. Rept. {\bf 319}, 145 (1999). 

\bibitem{Sof01}
S. Soff, S. A. Bass and A. Dumitru, Phys. Rev. Lett. {\bf 86}, 3981 (2001). 

\bibitem{Wei02}
R. M. Weiner, Phys. Rept. {\bf 327}, 249 (2002).

\bibitem{Pra03}
S. Pratt, Nucl. Phys. A {\bf 715}, 389c (2003).
                                                                                
\bibitem{Sof02}
S. Soff, S. A. Bass, D. H. Hardtke, and S. Y. Panitkin, J. Phys. G {\bf 28},
1885 (2002).
                                                                                
\bibitem{Hei02}
U. Heinz and P. Kolb, Nucl. Phys. A {\bf 702}, 269 (2002).
                                                                                
\bibitem{Lin02} Zi-Wei Lin, C. M. Ko, and Subrata Pal,
Phys. Rev. Lett. {\bf 89}, 152301 (2002).

\bibitem{Tea03}
D. Teaney, Nucl. Phys. A {\bf 715}, 817 (2003).
                                                                                
\bibitem{Cso03} T. Cs\" org\" o and J. Zim\' anyi, Acta
Phys. Hung. New Series, Heavy-Ion Physics, {\bf 17}, 281 (2003),
nucl-th/0206051.
                                                                                
\bibitem{Mol04}
D. Moln\'{a}r and M. Gyulassy, Phys. Rev. Lett. {\bf 92},
052301 (2004).
                                                                                
\bibitem{Zha04}
W. N. Zhang, M. J. Efaaf, C. Y. Wong, Phys. Rev. C {\bf 70}, 024903 (2004). 

\bibitem{Soc04} 
O. Socolowski Jr., F. Grassi, Y. Hama, and T. Kodama, 
Phys. Rev. Lett. {\bf 93}, 182301 (2004); 

\bibitem{Won03} C. Y. Wong, J. Phys. G {\bf 29}, 2151 (2003);
C. Y. Wong, J. Phys. G {\bf 30}, S1053 (2004); C. Y. Wong, AIP
Conference Proceedings, {\bf 828}, 617 (2006), hep-ph/0510258.

\bibitem{Cra05}
J. G. Cramer, G. A. Miller, J. M. S. Wu, and J. H. Yoon, Phys. Rev. Lett. 
{\bf 94}, 102302 (2005). 

\bibitem{Lis05} 
M. A. Lisa, S. Pratt, R. Soltz, U. Wiedemann, Ann. Rev. Nucl. Part. Sci. 55, 
357 (2005); nucl-ex/0505014.  

\bibitem{Pra05}
S. Pratt and D. Schindel, nucl-th/0511010. 

\bibitem{Fro06} 
E. Frodermann, U. Heinz, M. A. Lisa, Phys. Rev. C {\bf 73}, 044908 (2006).  

\bibitem{Hir02} 
T. Hirano and K. Tsuda, Phys. Rev. C {\bf 66}, 054905 (2002).  

\bibitem{Huo03} 
P. Huovinen, Nucl. Phys. A {\bf 715}, 299 (2003).  

\bibitem{Hum03} 
T. Humanic, Nucl. Phys. A {\bf 715}, 641 (2003); 
T. Humanic, Int. J. Mod Phys. E {\bf 15}, 197 (2006). 

\bibitem{Csa04} 
M. Csan\' ad, T. Cs\" org\' o, B. L\" rstad. and
A. Ster, J. Phys. G {\bf 30}, S1079 (2004); M. Csan\' ad, T. Cs\" org\'
o, B. L\" rstad, and A. Ster, nucl-th/0510027.

\bibitem{Lin05} 
Z. W. Lin, C. M. Ko, B. A. Li, B. Zhang, S. Pal, Phys. Rev. C {\bf 72}, 
064901 (2005).  

\bibitem{Tom05}
B. Tom\' a\u sik, nucl-th/0509100. 
                                      
\bibitem{Ham05}
Y. Hama, Rone P.G. Andrade, F. Grassi, O. Socolowski Jr, T. Kodama, 
B. Tavares, S. S. Padula,  hep-ph/0510096. 
                                      
\bibitem{Ris98} D. H. Rischke, Proceedings of the 11th Chris
Engelbrecht Summer School in Theoretical Physics, Cape Town, February
4-13, 1998, nucl-th/9809044.
             
\bibitem{Won04}
C. Y. Wong and W. N. Zhang,  Phys. Rev. C {\bf 70}, 064904 (2005). 

\bibitem{Zha05}
W. N. Zhang, S. X. Li, C. Y. Wong, and M. J. Efaaf, Phys. Rev. C {\bf 71}, 
064908 (2005). 

\bibitem{Wit84}
E. Witten, Phys. Rev. D {\bf 30}, 272 (1984).

\bibitem{Mye74}
W. D. Myers and W. J. Swiatecki, 
Ann. Phys. {\bf 84}, 186 (1974).

\bibitem{Mol93}
P. M\" oller, J. R. Nix, W. D. Myers, and W. J. Swiatecki,
Atom. Data Nucl. Data Tabl. {\bf 59}, 185  (1995). 

\bibitem{Kaj86}
K. Kajantie and H. Kurki-Suonio, Phys. Rev. D {\bf 34}, 1719 (1986).

\bibitem{Cse92}
L. P. Csernai and J. I. Kapusta,
Phys. Rev. Lett. {\bf 69}, 737 (1992).

\bibitem{Iwa92}
Y. Iwasaki $et~al.$, Phys. Rev. D {\bf 49}, 3540 (1994).

\bibitem{Gro93}
B. Grossmann and M. L. Lauren, Nucl. Phys. B {\bf 408}, 637 (1993). 

\bibitem{Bib97}
B. Beinlich, F. Karscha and A. Peikerta 
Phy. Lett. B {\bf 390}, 268 (1997). 

\bibitem{Kis05}
L. S. Kisslinger, S. Walawalkar, and M.B. Johnson,
Phys. Rev. D {\bf 71},  065017 (2005).

\bibitem{Dre02}
H. J. Drescher, F. M. Liu, S. Ostapchenko, T. Pierog, and K. Werner,
Phys. Rev. C {\bf 65}, 054902  (2002).

\bibitem{Lan53} L. D. Landau, Izv. Akad. Nauk. SSSR. ser. fiz. 
{\bf 17}, 51 (1953); L. D. Landau and S. Z. Belenkii, Usp. Fiz. Nauk 
{\bf 56}, 309 (1955).

\bibitem{Ste05}
P. Steinberg, Nucl. Phys. A {\bf 752},  423 (2005).

\bibitem{Mur04}
M. Murray, the BRAHMS Collaboration, J. Phys. G {\bf 30}, S667 (2004).

\bibitem{Won73}
C. Y. Wong, Ann. Phys. (N.Y.) {\bf 77}, 279 (1973).

\bibitem{Lan59}
L. D. Landau and E. M. Lifshitz, {\it Fluid Mechanics}, (Pergamon, New York, 
1959). 

\bibitem{Won76} C. Y.  Wong, J. Math. Phys. {\bf 17}, 1008 (1976);
C. Y. Wong, J. A. Maruhn, and T. A. Welton, Nucl. Phys. A {\bf 253},
469 (1975); C. Y. Wong, T. A. Welton, and J. A. Maruhn,
Phys. Rev. C {\bf 15}, 1558 (1977). 

\bibitem{Kar01}
F. Karsch, E. Laermann, and A. Peikert,
Nucl. Phys. B {\bf 605}, 579  (2001). 

\bibitem{Won77} C. Y. Wong, J. A. Maruhn, and T. A. Welton,
Phys. Lett. B {\bf 66}, 19 (1977).

\bibitem{Bla87}
J. P. Blaizot and J. Y. Ollitrault, Phys. Rev. D {\bf 36}, 916 (1987).
                                                                                
\bibitem{Lae96}
E. Laermann, Nucl. Phys. A {\bf 610}, 1 (1996).

\bibitem{PHE04b}
PHENIX Collaboration, S. S. Adler {\it et al.}, Phys. Rev. C {\bf 69}, 
034909, 2004.

\bibitem{Won82}
C. Y. Wong, Phys. Lett. B {\bf 88}, 39 (1979).

\bibitem{Oll92}
J. Y. Ollitrault, Phys. Rev. D {\bf 46}, 229 (1992). 

\bibitem{Vol96}
S. Voloshin, Y. Zhang, Z. Phys. Rev. C {\bf 70}, 665 (1996); 
hep-ph/9407282. 

\bibitem{Kol00}
P. F. Kolb, J. Sollfrank, U. Heinz, Phys. Rev. C {\bf 62}, 054909 (2000). 

\bibitem{Kol01}
P. F. Kolb, P. Huovinen, U. Heinz, H. Heiselberg, 
Phys. Lett. B {\bf 500}, 232, (2001). 

\bibitem{Huo01}
P. Huovinen, P. F. Kolb, U. Heinz, P. V. Ruuskanen, S. A. Voloshin, 
Phys. Lett. B {\bf 503}, 58, (2001). 

\bibitem{Kol03}
P. F. Kolb, R. Rapp, Phys. Rev. C {\bf 67}, 044903, (2003). 

\bibitem{Huo03a}
P. Huovinen, nucl-th/0305064.  

\bibitem{Sor05}
P. R. Sorensen, nucl-ex/0510052. 
                                      
\bibitem{Won94}
C. Y. Wong, {\it Introduction to High-Energy Heavy-Ion 
Collisions} (World Scientific, Singapore, 1994), Chap. 17. 

\bibitem{Pra90}
S. Pratt, Phys. Rev. Lett. {\bf 53}, 1219 (1984); S. Pratt, Phys. Rev. D 
{\bf 33}, 72 (1986);  S. Pratt, T. Cs\"{o}rgo, and J. Zim\'{a}nyi, Phys. 
Rev. C {\bf 42}, 2646 (1990). 

\bibitem{Ber88}
G. Bertsch, M. Gong, and M. Tohyama, Phys. Rev. C {\bf 37}, 1896 (1988); 
G. Bertsch, Nucl. Phys. A {\bf 498}, 173c (1989). 

\end{thebibliography}
\end{document}